\documentclass{aa}  
\usepackage{graphicx}
\usepackage{txfonts}
\newcommand{\AJ}[3]{{#1}, AJ, \vol{{#2}}, {#3}}
\newcommand{\ApJ}[3]{{#1}, ApJ, \vol{{#2}}, {#3}}
\newcommand{\ApJS}[3]{{#1}, ApJS, \vol{{#2}}, {#3}}

\newcommand{\AnnRev}[3]{{#1}, Annual Review of Astron. and Astrophys., \vol{{#2}}, {#3}}
\newcommand{\AandA}[3]{{#1}, A\&A, \vol{{#2}}, {#3}}
\newcommand{\AandAsupp}[3]{{#1}, A\&A Supp.\rm, \vol{{#2}}, {#3}}

\newcommand{\MNRAS}[3]{{#1}, MNRAS\rm, \vol{{#2}}, {#3}}

\newcommand{\PASP}[3]{{#1}, PASP\rm, \vol{{#2}}, {#3}}

\newcommand{\Acta}[3]{{#1}, Acta Astr.\rm, \vol{{#2}}, {#3}}
\newcommand{\vol}[1]{{\mbox{#1}}}

\newcommand{\Mv}{\mbox{$M_{V}\,$}}
\newcommand{\Mk}{\mbox{$M_{K}\,$}}
\newcommand{\Mj}{\mbox{$M_{J}\,$}}
\newcommand{\Mh}{\mbox{$M_{H}\,$}}
\newcommand{\Mi}{\mbox{$M_{I}\,$}}
\newcommand{\Wvi}{\mbox{$W_{VI}\,$}}
\newcommand{\Wjk}{\mbox{$W_{JK}\,$}}

\newcommand{\FeH}{\mbox{[Fe/H]}\,}

\newcommand{\kms}{\mbox{$\mbox{km\,s}^{-1}$}\,}

\newcommand{\magdex}{\mbox{mag~dex$^{-1}$}}

\begin{document}
   \title{Calibrating the Cepheid Period-Luminosity relation from the
infrared surface brightness technique}
\subtitle{II. The effect of metallicity, and the distance to the LMC\thanks{Based on observations collected at the 
European Organisation for Astronomical Research in the Southern Hemisphere, 
Chile, Programme-IDs 076-C.0158, 078.D-0299, \& 080.D-0318}$^,$\thanks{Table 2 
is only available in electronic form
at the CDS via anonymous ftp to cdsarc.u-strasbg.fr (130.79.128.5)
or via http://cdsweb.u-strasbg.fr/cgi-bin/qcat?J/A+A/}}

   \author{
          J. Storm\inst{1}
	  \and
	  W. Gieren\inst{2}
	  \and
	  P. Fouqu\'e\inst{3}
	  \and
          T.G. Barnes\inst{4}
	  \and
	  I. Soszy\'nski\inst{5}
	  \and
	  G. Pietrzy\'nski\inst{2,5}
	  \and
	  N. Nardetto\inst{2,6}
	  \and
	  D. Queloz\inst{7}
          }
%

   \institute{
             Leibniz-Institut f\"ur Astrophysik Potsdam (AIP),
             An der Sternwarte 16,
             D-14482 Potsdam, Germany,
             \email{jstorm@aip.de}
	 \and
	 Universidad de Concepci\'on, Departamento
         Astronom\'ia, Casilla 160-C, Concepci\'on, Chile
	 \and
         IRAP, Universit\'e de Toulouse, CNRS, 
         14 av. E. Belin, F-31400 Toulouse, France
	 \and
	 University of Texas at Austin, McDonald Observatory, 
         82  Mt.  Locke Rd, McDonald Observatory, TX 79734, USA
	\and
	Warsaw University Observatory, Al. Ujazdowskie 4, 00-478,
Warsaw, Poland
	\and
	Laboratoire Fizeau, UNS/OCA/CNRS UMR6525, Parc Valrose, 06108
Nice Cedex 2, France
        \and
	Observatoire Astronomique de l'Universit\'e de Gen\`eve, Chemin
de Maillettes 51, 1290 Sauverny, Switzerland
             }

   \date{Received 29 April 2011 / Accepted 23 July 2011}

\abstract
{
The extragalactic distance scale builds directly on the Cepheid
Period-Luminosity (PL) relation as delineated by the sample of Cepheids in
the Large Magellanic Cloud (LMC). However, the LMC is a dwarf irregular
galaxy, quite different from the massive spiral galaxies used for calibrating
the extragalactic distance scale. Recent investigations suggest that
not only the zero-point but also the slope of the Milky Way PL relation
differ significantly from that of the LMC, casting doubts on the universality
of the Cepheid PL relation.}
{
We want to make a differential comparison of the PL relations in the two
galaxies by delineating the PL relations using the same method, the 
infrared surface brightness method (IRSB), and the same precepts. 
}
{The IRSB method is a
Baade-Wesselink type method to determine individual distances to
Cepheids. We apply a newly revised  calibration of the method as described in an
accompanying paper (Paper I) to 36 LMC and five SMC Cepheids and
delineate new PL relations in the $V,I,J,$ \& $K$ bands as well as in
the Wesenheit indices in the optical and near-IR.}
{We present 509 new and accurate radial velocity measurements
for a sample of 22 LMC Cepheids, enlarging our earlier sample of 14 stars
to include 36 LMC Cepheids. The new calibration of the IRSB method is
directly tied to the recent HST parallax measurements to ten Milky Way
Cepheids, and we find a LMC barycenter distance modulus of 
$18.45\pm0.04$ (random error
only) from the 36 individual LMC Cepheid distances. In the $J,K$ bands
we find identical slopes for the LMC and Milky Way PL relations and only
a weak metallicity effect on the zero points (consistent with a zero effect), 
metal poor stars being fainter. In the optical we find the Milky Way slopes are slightly shallower
than the LMC slopes (but again consistent with no difference in the slopes)
 and small effects on the zero points. However, the
important Wesenheit index in $V,(V-I)$  shows a metallicity effect
on the slope and on the zero point which is likely to be significant.}
{We find a significant metallicity effect on the \Wvi index
$\gamma(\Wvi)=-0.23\pm0.10$~\magdex as well as an effect on the slope.
The $K$-band PL relation on the other hand is found to be an excellent extragalactic
standard candle being metallicity insensitive in both slope and 
zero-point and at the same time being reddening insensitive and showing the
least internal dispersion.}



\maketitle
%

\section{Introduction}
\label{sec.intro}

  The Cepheid Period-Luminosity (PL-) Relation is fundamental to the
calibration of the extra-galactic distance scale and thus to the
determination of the Hubble constant. 
Modern reviews on the calibration of the Cepheid distance scale can be 
found in e.g. Freedman and Madore (\cite{FM91}),
Fouqu\'e et al. (\cite{Fouque03}),  Sandage and Tammann (\cite{ST06}), 
Fouqu\'e et al.  (\cite{Fouque07}), and Barnes (\cite{Barnes09}) 
while Freedman and Madore (\cite{FM10}) review the present status of 
the quest for the Hubble constant. A dissenting view can be found e.g.
in Sandage et al. (\cite{Sandage09}) and references therein.

The value of the PL relation rests with its universality, in particular
that the PL relation slope and zero points are independent of metallicity.
The zero point has long been suggested to be metallicity dependent and
the HST Key Project on the Extragalactic Distance Scale (Freedman et al.
\cite{HSTkey}) corrected for this based on the empirical studies available 
at that time (e.g. Freedman and Madore \cite{FM90}, 
Sasselov et al. \cite{Sasselov97}, Kochanek \cite{Kochanek97},
Kennicutt et al. \cite{Kennicutt98}) with an estimated effect in the 
\Wvi index of $-0.2\pm0.2$~\magdex in the sense that metal-poor Cepheids 
are fainter than metal-rich Cepheids.  However, the size and even the 
sign of the effect is still disputed. Udalski et al. (\cite{Udalski01})
found no effect on the slope nor on the zero-point when comparing 
the metal-poor Cepheids of IC1613 with those in the Magellanic Clouds
based on TRGB distances to those galaxies. 
On the other hand Romaniello et al. (\cite{Romaniello08}) found a 
significant effect with the opposite sign when comparing Milky Way and 
Magellanic Cloud Cepheids with individual spectroscopic metallicity 
determinations.
Sakai et al. (\cite{Sakai04}) and Storm et al. (\cite{Storm04}), found an
effect of similar size and sign as that adopted by the HST key
project, but the latter study also showed a significant difference in the
slopes of the Milky Way and LMC relations, as also found by Sandage,
Tammann, and Reindl (\cite{STR04}). Recently Bono et al. (\cite{Bono10})
have argued for no significant effect in the \Wvi and \Wjk index on 
the basis of empirical as well as theoretical arguments.

Observationally, we are in the dilemma of either using a large sample of
Cepheids, like in the Large Magellanic Cloud (LMC), which constrains very well
the slope of the relation for a low metallicity sample of stars and which 
leaves the zero point to be determined from secondary indicators, 
or of using direct geometric distances (parallaxes) to a handful of 
nearby, solar metallicity, Milky Way Cepheids which constrain well 
the zero point of the PL relation but which do not constrain the slope 
very well.

Baade-Wesselink type methods which use the pulsational properties of
the Cepheids to determine direct distances to individual Cepheids
promise to resolve this dilemma by yielding direct individual
distances to a large sample of Milky Way and Magellanic Cloud Cepheids
spanning a significant range of metallicities. In particular, the
near-infrared surface brightness (IRSB) method (Fouqu\'e \& Gieren,
\cite{FG97}), a near-infrared variant of the Barnes-Evans method
(Barnes and Evans, \cite{BarnesEvans76}) shows great promise in
achieving this goal as it is insensitive to reddening errors. The
technique has also been shown to be independent of the metallicity
of the Cepheids (Storm et al. \cite{Storm04}). A few years ago
the IRSB method was re-calibrated using interferometrically
measured, phase-resolved angular diameters of Cepheids by Kervella et
al. (\cite{Kervella04b}). This was an extremely important achievement
as it proved that the pulsating Cepheid variables indeed obey the same
surface brightness-colour relation as the stable yellow giant stars
which were used to construct the surface brightness-colour relation
adopted in the original calibration of the technique (Fouqu\'e \&
Gieren, \cite{FG97}). The improved version of the IRSB technique was
then applied to Milky Way Cepheids by Storm et al. (\cite{Storm04})
and Groenewegen (\cite{Groenewegen08}), and for the first time to
extragalactic Cepheids by Storm et al. (\cite{Storm04}) (SMC) and
Gieren et al. (\cite{Gieren05a}) (LMC).

In our earlier application of the method to Milky Way Cepheids (Storm
et al. \cite{Storm04}), we found that the slopes of the resulting PL
relations in all optical and near-infrared bands were significantly
steeper than the those directly observed in the Large Magellanic
Cloud by the OGLE Project and by Persson et al. (\cite{Persson04}), 
thus challenging
the universality of the PL relation.  Gieren et al. (\cite{Gieren05a})
then analyzed thirteen LMC Cepheids, for which the data required for the
IRSB analysis were available at the time, in an identical fashion and
found that the PL relation slopes in the LMC were very similar to the
one obtained for the Milky Way sample, but different from the observed
apparent slope of the LMC sample.  We interpreted this as evidence for
the existence of an as yet undetected period-dependent systematic error
in the IRSB method, but the limited size of the LMC sample prevented
firm conclusions.

To put our previous analysis on a firmer basis, we present in the present
paper new and very accurate radial velocity curves for 22 additional
LMC Cepheids (Sec.\ref{sec.data}) thus almost tripling the existing
sample for an IRSB analysis to a total of 36 stars. To allow a purely
differential comparison with solar metallicity Cepheids we present a
similar analysis for 77 Milky Way Cepheids in an accompanying paper
(Storm et al. \cite{Storm11a}, hereinafter Paper I).  In that paper
we also combine the new empirical constraints on the projection
($p$-)factor, converting Cepheid radial into pulsational velocities,
as obtained from the present LMC study and the recent HST parallax
measurements to ten Galactic Cepheids (Benedict et al. \cite{Benedict07}),
to obtain a new empirical calibration of the $p$-factor relation 
to be used in our IRSB distance analysis.

In Sec.\ref{sec.IRSB} We briefly describe the IRSB method in its present
form and we discuss the choice of the $p$-factor relation. We present the
resulting LMC Period-Luminosity relations in optical and near-infrared
bands in Sec.\ref{sec.results} where we also compare to the SMC and Milky
Way samples to estimate the effect of metallicity on the PL relations.
In Sec.\ref{sec.discussion} 
we discuss our results and compare them to other recent
investigations. Sec.\ref{sec.summary}  summarizes our results.

\section{The Data}
\label{sec.data}

\subsection{The LMC Cepheid sample}

\begin{table}
\caption{\label{tab.new_sample}LMC Cepheids for which we have
obtained new radial velocity curves with the HARPS and FEROS
high-resolution spectrographs at ESO-La Silla.}
\begin{tabular}{r r r r}
\hline\hline
 HV873 & HV914 & HV2405 & HV12452 \\
 HV876 & HV1005 & HV2527 & HV12505 \\
 HV877 & HV1006 & HV2538 & HV12717 \\
 HV878 & HV1023 & HV2549 & U1 \\
 HV881 & HV2282 & HV5655 & \\
 HV900 & HV2369 & HV6093 & \\
\hline
\end{tabular}
\end{table}

\begin{figure}
\centering
\includegraphics[width=9cm]{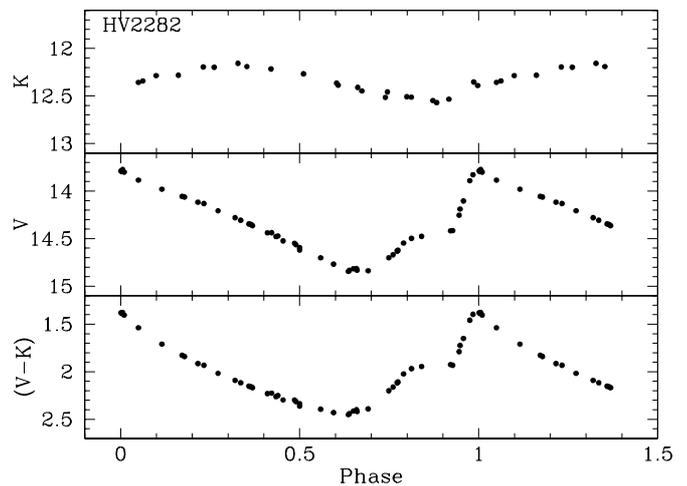}
\caption{\label{Fig.HV2282data} The photometric $V$ and $K$ light curves,
and the $(V-K)$ colour curve as used in our IRSB analysis for the 
LMC Cepheid variable
HV2282. This is a typical set of photometric light
and colour curves in the present study based on the optical photometry from 
OGLE-III (Udalski et al., \cite{Udalski08}, 
Soszy\'nski et al. \cite{Soszynski08}), and the
near-IR photometry from Persson et al. (\cite{Persson04}).}
\end{figure}

We have selected 22 Cepheids in the LMC (see Tab.\ref{tab.new_sample})
showing a wide range of periods, all having high-quality near-IR
lightcurves from Persson et al.  (\cite{Persson04}) and very accurate
optical photometry from OGLE-III (Udalski et al.  \cite{Udalski08},
Soszy\'nski et al. \cite{Soszynski08}). From their optical lightcurves,
all these Cepheids are clearly fundamental mode pulsators.  For these
stars we have obtained radial velocity curves as will be described
in the next section. A typical data set (HV~2282) is shown in
Fig.\ref{Fig.HV2282data}.

Gieren et al. (\cite{Gieren05a}) performed a first IRSB analysis on
a sample of 13 Cepheids in the LMC, six of which belonging to the
young blue, massive cluster NGC1866. These stars will all be used
in the following analysis.  To this sample we have added the star
\object{HV2827}. The radial velocity data for this star comes from
Imbert et al. (\cite{Imbert87}), the optical photometry from Moffet
et al. (\cite{Moffett98}) and the near-IR photometry from Persson et
al. (\cite{Persson04}).  For all the stars we have transformed the
near-IR photometry to the SAAO system using first the transformations
given by Persson et al.  (\cite{Persson04}) to transform the LCO data to
the CTIO system, and then applying the transformations given by Carter
(\cite{Carter90}) to transform from the CTIO to the SAAO system. In
this way we ensure to be on a common system with the Milky Way sample
presented in the accompanying Paper I, and to be on the same photometric
system which was used by Kervella et al.  (\cite{Kervella04b}) for the
calibration of the surface-brightness versus colour relation.

\subsection{Spectroscopic data}

  Using the HARPS instrument on the ESO 3.6m telescope and the FEROS
instrument on the ESO/MPG 2.2m telescope, both at ESO La Silla, Chile,
we have obtained 509 new radial velocity measurements for the above
described sample of 22 LMC Cepheids. HARPS was used for 20 of the stars
while FEROS was used for two additional Cepheids, HV914 and HV12717.

HARPS (Mayor et al. \cite{Mayor2003}) is a fiber-fed cross-dispersed echelle
spectrograph mounted inside a vacuum vessel for improved wavelength
stability.  It has a spectral resolution of $R=115000$ (3.2~pix) and the
wavelength range reaches from 380 to 690nm.  HARPS was used to observe 20
LMC Cepheids in the sample. Radial velocity observations were obtained
during a single, 16-night, visitor observing run in December 2005 which were
later complemented with service mode observations in the following two
seasons to cover phase gaps in the velocity curves.

We used the standard reduction pipeline using cross-correlation 
with a G2-type spectral mask (Baranne et al. \cite{Baranne96},
Pepe et al. \cite{Pepe02}) to extract radial velocities from the spectral
data. 

FEROS (Kaufer et al. \cite{Kaufer99}) is also a fiber-fed
crossed-dispersed spectrograph. 
It has a resolution of $R=48000$ (2.2~pix) covering the
wavelength range from 360 to 920nm. The Cepheid radial velocities were
extracted using the cross-correlation method with the FEROS pipeline.

Both instruments are known to be very stable with wavelength drifts
well below 100~m/s. As our main objective was to obtain well covered, accurate
radial velocity curves we made relatively short exposures with resulting
low signal to noise ratios of 5 to 10. This was found sufficient to keep 
the errors of the individual radial velocity measurements below 100~m/s, which
meets the precision we desired.

\begin{figure*}
\centering
\includegraphics[width=18cm]{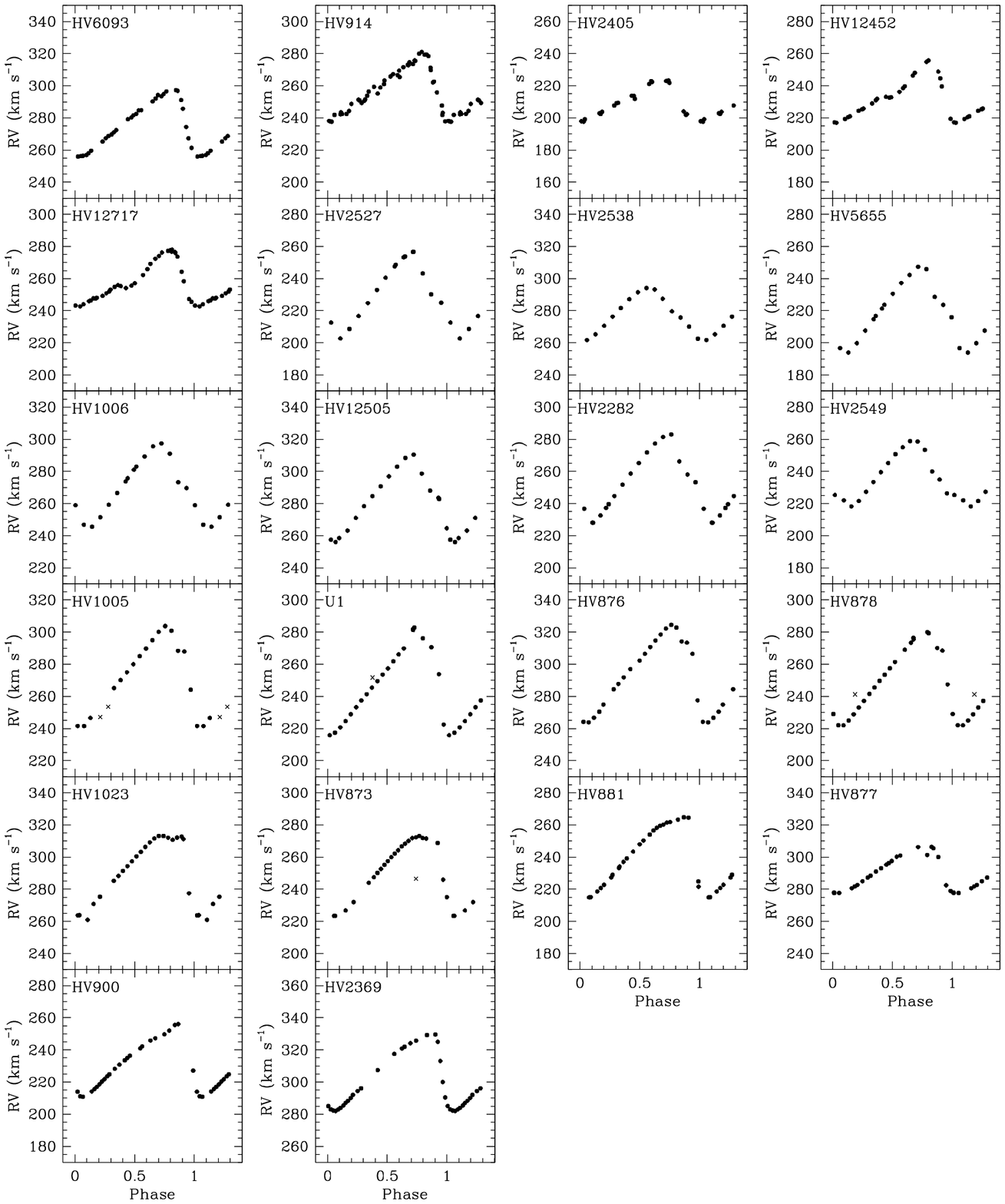}
\caption{\label{Fig.phrvall} The new radial velocity curves for 22
   LMC Cepheids from HARPS and FEROS (HV 914 and HV 12717) data. The
   crosses indicate points that have not been used in the analysis. The
   radial velocity range in each panel is 120~km~s$^{-1}$ so the amplitudes
   in the panels are directly comparable.  The panels are arranged
   according to increasing pulsational period.}
\end{figure*}

  The Heliocentric Julian dates and the individual radial velocities are
tabulated in Tab.\ref{tab.LMCrvdata} and the radial velocity curves are plotted
in Fig.\ref{Fig.phrvall}. A few data points appear to be
mis-identifications as they are close to the systemic velocity of the
star but do not fall on the velocity curve (see HV873, HV878, and U1). 
They have been marked with
crosses in the figure and are not considered in the further analysis. 
In the case of HV1005 we found two points with similar offsets from the
radial velocity curve. These two data points were obtained two years
later than the majority of the data, obtained during the sixteen 
consecutive nights (see above), thus suggesting the possibility of
orbital motion for this star. We have simply disregarded those two points 
in the following analysis.

\begin{table}
\caption{\label{tab.LMCrvdata} The new radial velocities for the 22
LMC Cepheids. The complete table is available in the electronic form
from the CDS.}
\begin{tabular}{r r r }
\hline\hline
\multicolumn{1}{c}{ID} & \multicolumn{1}{c}{HJD} & \multicolumn{1}{c}{RV} \\
\multicolumn{1}{c}{} & \multicolumn{1}{c}{(days)} &
\multicolumn{1}{c}{(\kms)} \\
\hline
   HV873 & 2453701.7205 & 247.46 \\
   HV873 & 2453702.7489 & 250.14 \\
   HV873 & 2453703.7567 & 252.64 \\
   HV873 & 2453704.7536 & 255.15 \\
   HV873 & 2453705.7469 & 257.52 \\
   HV873 & 2453706.7521 & 259.96 \\
     ... & ... & ... \\
\hline
\end{tabular}
\end{table}

\section{The IRSB method}
\label{sec.IRSB}

\subsection{The surface brightness-colour relation}

The infrared surface brightness (IRSB) method is a variant of the
Baade-Wesselink method originally developed by Barnes and Evans
(\cite{BarnesEvans76}) in the optical spectral range. It matches the
angular diameter variation of a Cepheid as determined from photometry,
with the radius variation of the pulsating star as determined from an
integration of its pulsational velocity curve.

In Paper I we present a more detailed discussion of the method, which
relates the surface brightness parameter, $F_V$, to a colour index
$(V-K)_0$ to determine the angular diameter variation of the star.

Here we use the relation

\begin{eqnarray}
F_V & = & 4.2207 - 0.1 V_0 - 0.5 \log \theta\\
    & = & -0.1336 (V-K)_0 + 3.9530
\end{eqnarray}
\noindent
as determined by Kervella et al. (\cite{Kervella04b}).

\subsection{The projection factor}
\label{sec.pfactor}

Knowledge of the projection ($p$-) factor which converts an observed
radial velocity of a Cepheid into the pulsational velocity at its
surface is crucial for any kind of Baade-Wesselink
analysis.  The $p$-factor has to take into account not only the
geometrical projection effects across the observed stellar disk but
also fold this with the effect of limb darkening and possibly
take into consideration non-LTE effects due to the dynamic behaviour 
of the pulsating atmosphere.

Recent empirical (Gieren et al., \cite{Gieren05a}) and theoretical
(Nardetto et al., \cite{Nardetto07} \& \cite{Nardetto09}) work has shown
that the $p$-factor relation from Hindsley and Bell (\cite{Hindsley89}),
which shows a slight period dependence of the $p$-factor and was used in
our early work (Gieren et al. 1993), is not appropriate. In fact the new
data presented here (see Sec.\ref{sec.pfac_determ}) largely supports
the empirical findings of Gieren et al. (\cite{Gieren05a}) that the
$p$-factor depends quite strongly on pulsational period or the derived
distances to the LMC Cepheids become period dependent, which is clearly
unphysical. We have discussed this issue in more detail in Paper I. In that
paper, we find the best fitting relation to be: 
\begin{equation}
p = 1.550(\pm0.04) - 0.186(\pm0.06) \log P
\end{equation}
\noindent
which we will adopt here. 

It is important to note that the exact choice of the $p$-factor relation
has no bearing on the conclusions regarding the effect of metallicity on
the PL relations in the present paper, as long as the $p$-factor relation
used in the analysis is the same for both samples, the Milky Way and
the LMC. Any change in the relation would affect both Milky Way and
LMC Cepheids in the same way but the differentials would remain the same.
This of course is based on the implicit assumption that the $p$-factor
is independent of metallicity. Work is currently underway to further
investigate this point.

\begin{figure}
\centering
\includegraphics[width=9cm]{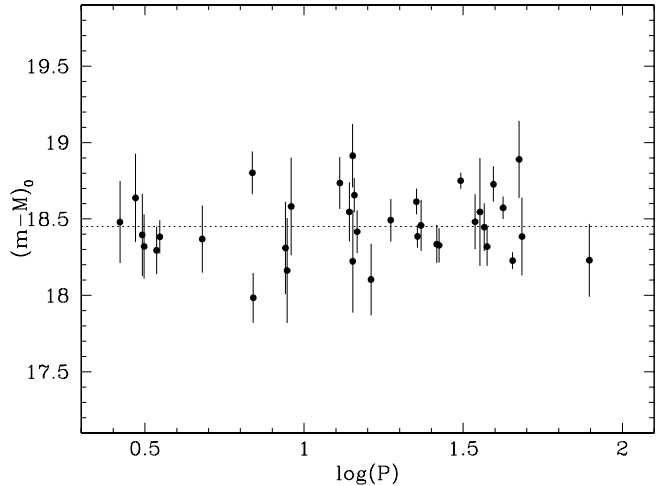}
\caption{\label{Fig.logPmM} The IRSB-based distance moduli of 36 Cepheids in the LMC
using the new $p$-factor relation derived in Paper I,
plotted as a function of their pulsation period. The moduli have been
 referred to the LMC barycenter using the LMC disk model of van der Marel and
 Cioni (\cite{Marel01}).}
\end{figure}

\subsection{The reddening}

  The reddenings for the LMC stars were taken from the catalogue of Persson
et al. (\cite{Persson04}) whenever possible. For the NGC1866 stars 
we adopted values of $E(B-V)=0.06$ following the discussion in 
Storm et al. (\cite{Storm05}).
The reddening is very low for all the stars and the star-to-star variation is very
low as well, so reddening errors can only marginally affect the resulting
Period-Luminosity relations, even in the optical bands. We have adopted
the reddening law from Cardelli et al. (\cite{Cardelli89}) with a
total-to-selective absorption of $R_V=3.23$ following the
discussion in Fouqu\'e et al. (\cite{Fouque07}). In this way we are
proceeding exactly in the same way as for the Milky Way Cepheids in Paper I.

\subsection{Adopted metallicity}
\label{sec.abundance}

  We adopt the mean metallicity estimates for the Milky Way, LMC and 
SMC Cepheids from
the discussion by Luck et al. (\cite{Luck98}) based on new and literature 
measures for Cepheids and supergiants. They find for the Milky Way
$\FeH=+0.03$ based on 69 stars with $\sigma=0.14$, for the LMC
$\FeH=-0.34$ based on 32 stars with $\sigma=0.15$ and for the SMC
$\FeH=-0.68$ based on 25 stars with $\sigma=0.13$.
These values are largely supported by Romaniello et al.
(\cite{Romaniello08}) who found values of $-0.30$~dex and $-0.75$~dex
respectively.
We note that the Cepheids in NGC1866 might be slightly more metal-poor,
$\FeH=-0.5$, based on the spectroscopic study of three cluster giants
by Hill et al. (\cite{Hill00}) and the photometric measurement for
NGC1866 of $-0.48\pm0.18$ by Hilker et al.  (\cite{Hilker95}) from
Str\"omgren photometry, but we do not make this distinction in the following. 

\clearpage

\section{Results}
\label{sec.results}

\begin{table*}
\caption{\label{tab.LMCresults} Distances and absolute magnitudes
for the LMC Cepheids calculated using the precepts given in 
Paper I, using the new $p$-factor relation. In columns 3 and 4 we
give the IRSB distance and the associated error estimate. In
columns 5 and 6 follows the distance modulus, 
and in the columns 7 through 13, we give the absolute magnitudes
in the different bands as well as the Wesenheits indices. Finally in
column 14 we give the adopted reddening and in column 15 the adopted
phase shift between spectroscopic and photometric angular diameters. Column 16
gives the magnitude correction to refer the distance moduli to the LMC barycenter.}
\scriptsize
\begin{tabular}{r r r r r r r r r r r r r r r r}
\hline\hline
(1) & (2) & (3) & (4) & (5) & (6) & (7) & (8) & (9) & (10) & (11) & (12)
& (13) & (14) & (15) & (16) \\
ID & $\log (P)$ & $d$ & $\sigma(d)$ & $(m-M)_0$ & $\sigma_{(m-M)}$ & $\Mv$ & $\Mi$ & $\Mj$ & $\Mh$ & $\Mk$ & $\Wvi$ & $\Wjk$ & $E(B-V)$ & $\Delta \phi$ & $\Delta (m-M)$\\
 &  & (kpc) & (kpc) & (mag) & (mag) & (mag) & (mag) & (mag) & (mag) &
(mag) & (mag) & (mag) & (mag) & & (mag) \\
\hline
\object{   HV 12199} &   0.421469 &  49.7 &  1.8 & 18.48 & 0.08 & $-2.41$ & $-3.01$ & $-3.13$ & $$ & $-3.80$ & $-3.94$ & $-4.26$ & $ 0.060$ & $ 0.035$ & $-0.058$\\
\object{   HV 12203} &   0.470427 &  53.4 &  2.1 & 18.64 & 0.09 & $-2.71$ & $-3.31$ & $-3.66$ & $$ & $-4.08$ & $-4.24$ & $-4.37$ & $ 0.060$ & $ 0.050$ & $-0.058$\\
\object{   HV 12202} &   0.491519 &  47.8 &  1.7 & 18.40 & 0.08 & $-2.53$ & $-3.14$ & $-3.60$ & $$ & $-4.02$ & $-4.09$ & $-4.30$ & $ 0.060$ & $ 0.030$ & $-0.058$\\
\object{   HV 12197} &   0.497456 &  46.1 &  1.3 & 18.32 & 0.06 & $-2.42$ & $-3.07$ & $-3.44$ & $$ & $-3.87$ & $-4.07$ & $-4.17$ & $ 0.060$ & $-0.015$ & $-0.057$\\
\object{   HV 12204} &   0.536402 &  45.6 &  1.0 & 18.30 & 0.05 & $-2.79$ & $-3.33$ & $-3.67$ & $$ & $-4.07$ & $-4.17$ & $-4.34$ & $ 0.060$ & $ 0.015$ & $-0.059$\\
\object{   HV 12198} &   0.546887 &  47.5 &  0.7 & 18.38 & 0.03 & $-2.63$ & $-3.27$ & $-3.69$ & $$ & $-4.09$ & $-4.26$ & $-4.36$ & $ 0.060$ & $ 0.020$ & $-0.059$\\
\object{    HV 6093} &   0.679885 &  47.2 &  1.4 & 18.37 & 0.06 & $-3.25$ & $-3.86$ & $-4.21$ & $$ & $-4.60$ & $-4.79$ & $-4.87$ & $ 0.058$ & $-0.015$ & $-0.048$\\
\object{     HV 914} &   0.837489 &  57.6 &  1.1 & 18.80 & 0.04 & $-4.10$ & $-4.79$ & $-5.18$ & $-5.54$ & $-5.59$ & $-5.84$ & $-5.88$ & $ 0.070$ & $-0.035$ & $ 0.013$\\
\object{    HV 2405} &   0.840331 &  39.5 &  0.9 & 17.98 & 0.05 & $-3.06$ & $-3.77$ & $-4.14$ & $-4.52$ & $-4.57$ & $-4.86$ & $-4.86$ & $ 0.070$ & $-0.015$ & $ 0.046$\\
\object{   HV 12452} &   0.941457 &  45.9 &  1.9 & 18.31 & 0.09 & $-3.74$ & $-4.48$ & $-4.94$ & $-5.32$ & $-5.37$ & $-5.62$ & $-5.67$ & $ 0.058$ & $ 0.030$ & $ 0.058$\\
\object{   HV 12717} &   0.946628 &  42.9 &  2.0 & 18.16 & 0.10 & $-3.62$ & $-4.33$ & $-4.77$ & $-5.14$ & $-5.21$ & $-5.42$ & $-5.51$ & $ 0.058$ & $-0.005$ & $ 0.065$\\
\object{   HV 12816} &   0.959466 &  52.0 &  2.3 & 18.58 & 0.09 & $-4.29$ & $-4.88$ & $-5.24$ & $-5.56$ & $-5.62$ & $-5.79$ & $-5.88$ & $ 0.070$ & $ 0.015$ & $-0.075$\\
\object{    HV 2527} &   1.112251 &  55.8 &  1.3 & 18.73 & 0.05 & $-4.36$ & $-5.21$ & $-5.66$ & $-6.09$ & $-6.15$ & $-6.51$ & $-6.49$ & $ 0.070$ & $ 0.005$ & $ 0.041$\\
\object{    HV 2538} &   1.142118 &  51.2 &  1.3 & 18.55 & 0.06 & $-4.39$ & $-5.28$ & $-5.70$ & $-6.14$ & $-6.19$ & $-6.64$ & $-6.53$ & $ 0.100$ & $-0.040$ & $-0.019$\\
\object{    HV 5655} &   1.152657 &  60.7 &  1.7 & 18.91 & 0.06 & $-4.69$ & $-5.53$ & $-5.96$ & $-6.41$ & $-6.47$ & $-6.84$ & $-6.81$ & $ 0.100$ & $-0.015$ & $ 0.046$\\
\object{    HV 1006} &   1.152786 &  44.1 &  2.0 & 18.22 & 0.10 & $-4.17$ & $-4.95$ & $-5.48$ & $-5.90$ & $-5.95$ & $-6.15$ & $-6.28$ & $ 0.100$ & $ 0.005$ & $-0.010$\\
\object{   HV 12505} &   1.158115 &  53.9 &  0.8 & 18.66 & 0.03 & $-4.14$ & $-5.04$ & $-5.63$ & $-6.11$ & $-6.17$ & $-6.43$ & $-6.55$ & $ 0.100$ & $-0.005$ & $ 0.031$\\
\object{    HV 2282} &   1.166636 &  48.2 &  0.9 & 18.42 & 0.04 & $-4.41$ & $-5.19$ & $-5.66$ & $-6.07$ & $-6.12$ & $-6.38$ & $-6.44$ & $ 0.100$ & $ 0.015$ & $ 0.051$\\
\object{    HV 2549} &   1.210289 &  41.8 &  1.3 & 18.10 & 0.07 & $-4.50$ & $-5.31$ & $-5.69$ & $-6.06$ & $-6.13$ & $-6.56$ & $-6.43$ & $ 0.058$ & $ 0.020$ & $ 0.052$\\
\object{    HV 1005} &   1.272209 &  49.9 &  0.9 & 18.49 & 0.04 & $-4.72$ & $-5.47$ & $-5.98$ & $-6.41$ & $-6.45$ & $-6.62$ & $-6.77$ & $ 0.100$ & $ 0.015$ & $-0.023$\\
\object{        U1} &   1.353098 &  52.8 &  0.6 & 18.61 & 0.02 & $-4.82$ & $-5.70$ & $-6.28$ & $-6.74$ & $-6.80$ & $-7.07$ & $-7.16$ & $ 0.100$ & $ 0.010$ & $ 0.060$\\
\object{     HV 876} &   1.356342 &  47.6 &  0.5 & 18.39 & 0.02 & $-5.06$ & $-5.86$ & $-6.27$ & $-6.67$ & $-6.71$ & $-7.10$ & $-7.02$ & $ 0.100$ & $ 0.005$ & $ 0.005$\\
\object{     HV 878} &   1.367445 &  49.2 &  1.1 & 18.46 & 0.05 & $-5.11$ & $-5.66$ & $-6.35$ & $-6.75$ & $-6.80$ & $-6.52$ & $-7.12$ & $ 0.058$ & $-0.010$ & $ 0.055$\\
\object{   HV 12815} &   1.416910 &  46.5 &  0.8 & 18.34 & 0.04 & $-5.08$ & $-5.96$ & $-6.48$ & $-6.94$ & $-6.99$ & $-7.32$ & $-7.35$ & $ 0.070$ & $-0.030$ & $-0.075$\\
\object{    HV 1023} &   1.424235 &  46.3 &  0.7 & 18.33 & 0.03 & $-4.77$ & $     $ & $-6.28$ & $-6.75$ & $-6.79$ & $     $ & $-7.15$ & $ 0.070$ & $ 0.025$ & $-0.049$\\
\object{     HV 899} &   1.492039 &  56.2 &  0.4 & 18.75 & 0.01 & $-5.71$ & $-6.52$ & $-7.02$ & $-7.45$ & $-7.50$ & $-7.78$ & $-7.82$ & $ 0.110$ & $ 0.025$ & $ 0.018$\\
\object{     HV 873} &   1.537311 &  49.7 &  1.2 & 18.48 & 0.05 & $-5.87$ & $     $ & $-7.08$ & $-7.46$ & $-7.50$ & $     $ & $-7.79$ & $ 0.130$ & $-0.010$ & $ 0.080$\\
\object{     HV 881} &   1.552820 &  51.2 &  2.5 & 18.55 & 0.10 & $-5.55$ & $     $ & $-6.97$ & $-7.41$ & $-7.50$ & $     $ & $-7.85$ & $ 0.030$ & $-0.065$ & $ 0.062$\\
\object{     HV 879} &   1.566167 &  48.9 &  1.0 & 18.45 & 0.05 & $-5.27$ & $-6.25$ & $-6.84$ & $-7.32$ & $-7.39$ & $-7.75$ & $-7.78$ & $ 0.060$ & $ 0.015$ & $ 0.044$\\
\object{     HV 909} &   1.574986 &  46.1 &  0.8 & 18.32 & 0.04 & $-5.75$ & $-6.52$ & $-6.97$ & $-7.36$ & $-7.43$ & $-7.70$ & $-7.74$ & $ 0.058$ & $-0.065$ & $ 0.048$\\
\object{    HV 2257} &   1.595153 &  55.6 &  0.8 & 18.73 & 0.03 & $-5.90$ & $-6.80$ & $-7.35$ & $-7.81$ & $-7.87$ & $-8.19$ & $-8.22$ & $ 0.060$ & $-0.005$ & $ 0.054$\\
\object{    HV 2338} &   1.625350 &  51.8 &  0.5 & 18.57 & 0.02 & $-5.95$ & $-6.83$ & $-7.36$ & $-7.80$ & $-7.86$ & $-8.19$ & $-8.21$ & $ 0.040$ & $-0.015$ & $ 0.070$\\
\object{     HV 877} &   1.655215 &  44.2 &  0.3 & 18.23 & 0.02 & $-5.21$ & $     $ & $-6.80$ & $-7.31$ & $-7.36$ & $     $ & $-7.74$ & $ 0.100$ & $ 0.070$ & $ 0.017$\\
\object{     HV 900} &   1.675637 &  60.0 &  2.0 & 18.89 & 0.07 & $-6.32$ & $     $ & $-7.72$ & $-8.17$ & $-8.25$ & $     $ & $-8.62$ & $ 0.058$ & $-0.060$ & $ 0.042$\\
\object{    HV 2369} &   1.684646 &  47.5 &  1.6 & 18.39 & 0.07 & $-6.02$ & $     $ & $-7.46$ & $-7.90$ & $-7.96$ & $     $ & $-8.29$ & $ 0.095$ & $ 0.005$ & $-0.029$\\
\object{    HV 2827} &   1.896354 &  44.3 &  1.4 & 18.23 & 0.07 & $-6.18$ & $-7.20$ & $-7.81$ & $-8.29$ & $-8.36$ & $-8.75$ & $-8.73$ & $ 0.080$ & $ 0.035$ & $-0.079$\\
\hline
\end{tabular}
\end{table*}

\begin{table*}
\caption{\label{tab.SMCresults} Distances and absolute magnitudes
for the SMC Cepheids calculated by using the same precepts as in Table 3. 
Column headings are the same as in the previous table.}
\scriptsize
\begin{tabular}{r r r r r r r r r r r r r r r}
\hline\hline
ID & $\log P$ & $d$ & $\sigma(d)$ & $(m-M)_0$ & $\sigma_{(m-M)}$ & $\Mv$ & $\Mi$ & $\Mj$ & $\Mk$ & $\Wvi$ & $\Wjk$ & $E(B-V)$ & $\Delta \phi$ \\
 &  & (kpc) & (kpc) & (mag) & (mag) & (mag) & (mag) & (mag) & (mag) &
(mag) & (mag) & (mag) & \\
\hline
\object{    HV 1345} &   1.129638 &  55.5 &  1.8 & 18.72 & 0.07 &  $-4.06$ & $-4.80$ & $-5.41$ & $-5.88$ & $-5.96$ & $-6.20$ & $ 0.030$ & $ 0.025$\\
\object{    HV 1335} &   1.157807 &  59.9 &  1.6 & 18.89 & 0.06 &  $-4.37$ & $-5.06$ & $-5.56$ & $-5.95$ & $-6.12$ & $-6.21$ & $ 0.090$ & $-0.055$\\
\object{    HV 1328} &   1.199645 &  52.2 &  2.1 & 18.59 & 0.09 &  $-4.47$ & $-5.15$ & $-5.60$ & $-6.00$ & $-6.20$ & $-6.28$ & $ 0.000$ & $ 0.020$\\
\object{    HV 1333} &   1.212014 &  76.5 &  1.4 & 19.42 & 0.04 & $-4.94$ & $-5.69$ & $-6.20$ & $-6.63$ & $-6.84$ & $-6.92$ & $ 0.070$ & $-0.005$\\
\object{     HV 822} &   1.223810 &  62.1 &  2.2 & 18.96 & 0.08 & $-4.54$ & $-5.38$ & $-5.89$ & $-6.33$ & $-6.67$ & $-6.64$ & $ 0.030$ & $-0.010$\\
\hline
\end{tabular}
\end{table*}

\subsection{Distances and absolute magnitudes}

We have applied the IRSB method as described in the previous section
to find the distances and absolute magnitudes for the LMC Cepheids reported in
Tab.\ref{tab.LMCresults} and for the SMC sample of Storm et al.
(\cite{Storm04}) in Tab.\ref{tab.SMCresults}. The absolute magnitudes
are intensity-averaged magnitudes.

\subsection{The distances to the LMC and the SMC}

On the basis of the individual distances to the Cepheids in the LMC we
can now determine the distance to the LMC barycenter. First we correct
the measured distance of each Cepheid by the distance offset from the
LMC disk model of van der Marel \& Cioni (\cite{Marel01}) as tabulated
in the last column of Tab.\ref{tab.LMCresults} and then compute the mean
of the resulting values.  We find $(m-M)_0(LMC)=18.45\pm0.04$ (random
error only) with a standard deviation of 0.22~mag.  As the Cepheids are
well distributed across the face of the LMC the uncorrected mean modulus
is identical to within 0.01~mag to the adopted barycentric distance 
modulus value.

  For the SMC we only have five stars so the random error is
significantly larger than is the case for the LMC. Additionally, the 
SMC is well known to exhibit more pronounced depth effects than the LMC, 
so some additional scatter is to be expected. We find from the five Cepheids
an unweighted mean value of $(m-M)_0(\mbox{SMC})=18.92\pm0.14$ with a
standard deviation of 0.32~mag. The offset in modulus with
respect to the LMC of 0.47~mag is in excellent agreement with the very
accurate value of $0.44\pm0.05$~mag found by Cioni et al.
(\cite{Cioni00}) from an investigation of a large sample 
of tip of the red-giant branch (TRGB) stars in the reddening 
insensitive near-infrared, thus giving us confidence that our five stars 
are indeed representative of the SMC as a whole.

\subsection{Constraints on the $p$-factor relation}
\label{sec.pfac_determ}

\begin{table}
\caption{\label{tab.pfac} The derived LMC true distance modulus and the
slope of the individual Cepheid distance moduli as a function of their pulsation period,
for different $p$-factor relations (see text) $p=\beta_p + \alpha_p \log P$ }
\begin{tabular}{l l l l }
\hline\hline
$\alpha_{p}$ & $\beta_{p}$ & $(m-M)_0$ & $(m-M)_0$ slope \\
\hline
$-0.03$  & 1.39  & $18.50\pm0.04$ & $0.31\pm0.10$ \\
$-0.08$  & 1.31  & $18.26\pm0.04$ & $0.22\pm0.10$ \\
$-0.08$  & 1.455 & $18.50\pm0.04$ & $0.24\pm0.10$ \\
$-0.186$ & 1.550 & $18.45\pm0.04$ & $0.00\pm0.10$ \\
\hline
\end{tabular}
\end{table}

As discussed in Sec.\ref{sec.pfactor} we can use the IRSB distances to the
36 LMC Cepheids to place constraints on the $p$-factor relation.  We have
rederived the distances to the individual LMC Cepheids using different
$p$-factor relations, as discussed in more detail in Paper I,
and we computed the individual LMC barycentric distances as before.
In Tab.\ref{tab.pfac} we list the resulting mean LMC distance moduli
as well as the slope of the distance modulus as a function of pulsation
period, for each of the $p$-factor relations we tested in this process. The
function distance modulus versus pulsation period must of course have a
zero slope, and with the present large sample of 36 stars we are able to
place strong constrains on the slope. In the table it can be seen that
the original Hindsley and Bell (\cite{Hindsley89}) relation gives a very
plausible LMC distance but a strong and unphysical period dependence of
the distance moduli. Similarly the Nardetto et al. (\cite{Nardetto09})
relation gives a slope which deviates at the level of 2$\sigma$ from
the uncorrelated relation and the LMC distance becomes uncomfortably
short at 18.26~mag. The third relation uses the slope of the $p$-factor
relation from Nardetto et al.  (\cite{Nardetto09}) but with a zero point
which yields IRSB distances to Milky Way Cepheids in agreement with
the recent parallaxes from Benedict et al. (\cite{Benedict07}). The
last column corresponds to the relation derived in Paper~I which
we have adopted here, which yields LMC Cepheid distance moduli which
are totally independent of their pulsation period.  This is borne out
in Fig.\ref{Fig.logPmM} where we show the  LMC center distance moduli
based on the individual LMC Cepheid moduli corrected for the disk model
of van der Marel and Cioni (\cite{Marel01}) as a function of $\log
P$. It is evident that with the adopted $p$-factor relation there is no
significant correlation between pulsational period and derived distance
to the LMC Cepheids.

\subsection{The LMC Period-Luminosity relations}

\begin{figure}
\centering
\includegraphics[width=9cm]{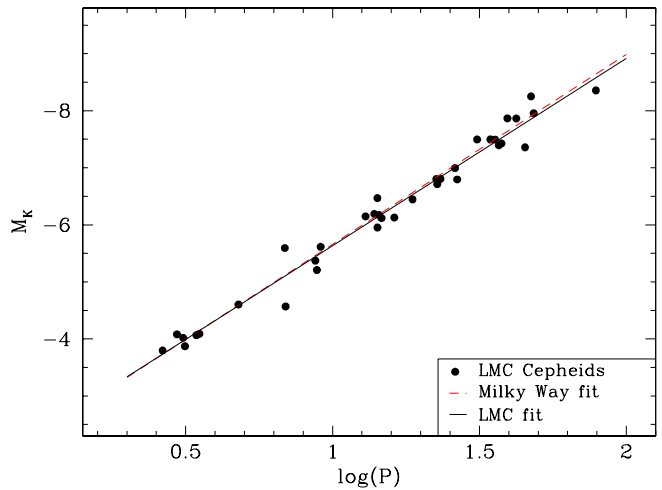}
\caption{\label{Fig.logPMk} The $K$-band period-luminosity relation for
our sample of LMC Cepheids based on the absolute magnitudes determined
with the IRSB method as calibrated in Paper I. The best fitting line
from the data is overplotted in black, and the best fitting line to
the Milky Way sample of Paper I is overplotted with a dashed line in
red. It is appreciated that the LMC and Milky Way PL relations agree
extremely well both in slope and zero point.}
\end{figure}

\begin{figure}
\centering
\includegraphics[width=9cm]{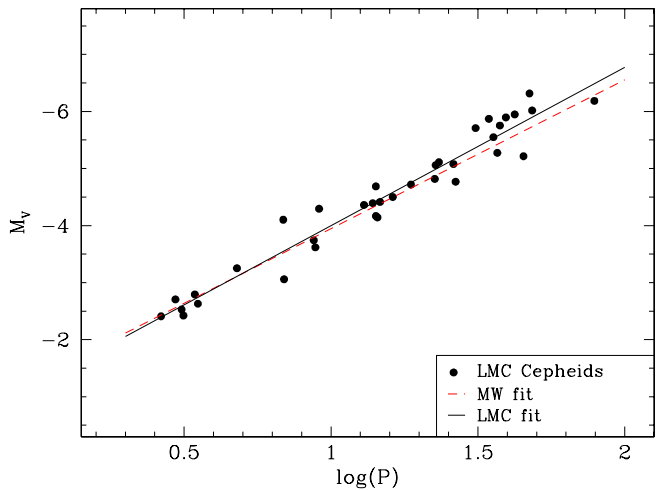}
\caption{\label{Fig.logPMv} The $V$-band PL relation
 for our sample of LMC Cepheids based on the absolute magnitudes
 determined by the IRSB method. Overplotted is the linear regression
fit (solid line) as well as the corresponding Milky Way relation (dashed line).}
\end{figure}

\begin{figure}
\centering
\includegraphics[width=9cm]{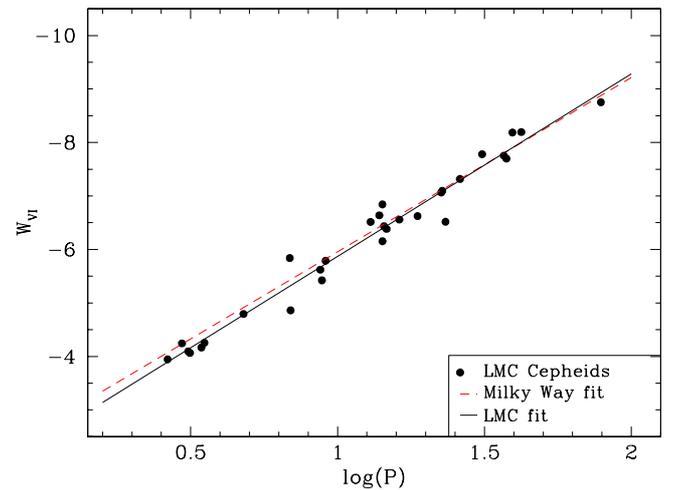}
\caption{\label{Fig.logPW} The optical $(V-I)$ Wesenheit index PL relation
 for our sample of LMC Cepheids, based on the absolute magnitudes
 determined by the IRSB method as calibrated in Paper I. 
 Overplotted is the linear regression fit (solid line) as well as the
corresponding Milky Way relation (dashed line).}
\end{figure}

\begin{figure}
\centering
\includegraphics[width=9cm]{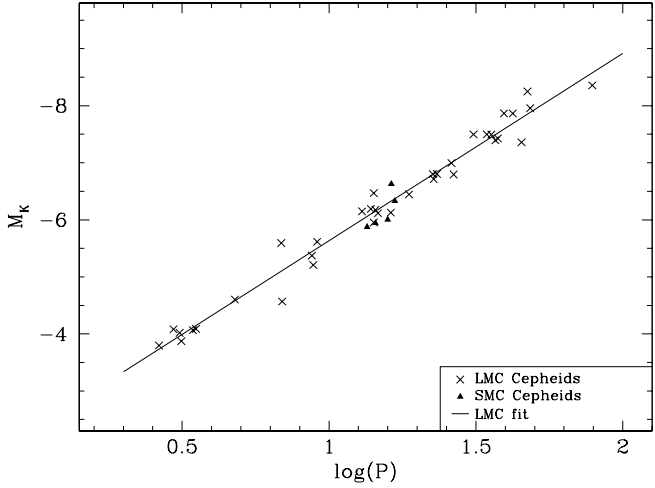}
\caption{\label{Fig.logPMkMC} The $K$-band absolute magnitudes
for our sample of SMC Cepheids (filled triangles) overplotted on the
LMC Cepheid PL relation (Fig.\ref{Fig.logPMk}), crosses are the individual
LMC Cepheids.}
\end{figure}

\begin{table}
\caption{\label{tab.PLrelations}The Period-Luminosity relations for the LMC
in the various optical and near-infrared bands
as determined from a linear regression to the absolute
magnitudes from the IRSB analysis. The relations have been fitted in the
form $M = a [\log(P)-1] + b$.  The fourth column gives the
dispersion around the fit. Our results for the Milky Way Cepheids are
included for comparison.}

\begin{tabular}{r c c c }
\hline\hline
\multicolumn{1}{c}{Band} & \multicolumn{1}{c}{$a$} &
\multicolumn{1}{c}{$b$} & \multicolumn{1}{c}{dispersion} \\
  & \multicolumn{1}{c}{(\magdex)} & \multicolumn{1}{c}{(mag)} & 
\multicolumn{1}{c}{(mag)} \\
\hline
\Mv(LMC) & $-2.78\pm0.11$ & $-4.00\pm0.05$ & 0.26 \\
\Mv(MW)  & $-2.67\pm0.10$ & $-3.96\pm0.03$ & 0.26 \\
 & & & \\
\Mi(LMC) & $-3.02\pm0.10$ & $-4.74\pm0.04$ & 0.21 \\
\Mi(MW)  & $-2.81\pm0.10$ & $-4.76\pm0.03$ & 0.23 \\
 & & & \\
\Wvi(LMC) & $-3.41\pm0.11$  & $-5.87\pm0.05$ & 0.24 \\
\Wvi(MW)  & $-3.26\pm0.11$  & $-5.96\pm0.04$ & 0.26 \\
 & & & \\
\Mj(LMC) & $-3.22\pm0.09$ & $-5.17\pm0.04$ & 0.21 \\
\Mj(MW)  & $-3.18\pm0.09$ & $-5.22\pm0.03$ & 0.22 \\
 & & & \\
\Mk(LMC) & $-3.28\pm0.09$ & $-5.64\pm0.04$ & 0.21 \\
\Mk(MW)  & $-3.33\pm0.09$ & $-5.66\pm0.03$ & 0.22 \\
 & & & \\
\Wjk(LMC) & $-3.31\pm0.09$  & $-5.95\pm0.04$ & 0.21 \\
\Wjk(MW)  & $-3.44\pm0.09$  & $-5.96\pm0.03$ & 0.23 \\
\hline
\end{tabular}
\end{table}

  In Tab.\ref{tab.PLrelations} the Period-Luminosity relations we obtain
for the LMC in the different photometric bands are given in the form:

\begin{equation}
\label{eq.logPM}
M = a (\pm \sigma(a)) [\log (P) - 1.0]  + b (\pm \sigma(b))
\end{equation}

The relations have been determined from a linear regression to the
absolute magnitudes and $\log (P)$ values given in Tab.\ref{tab.LMCresults}.
In the table the corresponding relations for the Milky Way sample
determined in Paper I are given for comparison. These relations
are based on Cepheid distances calculated with exactly the same precepts
as in this paper, using the same $p$-factor relation and IRSB calibration
so we can perform a direct comparison of the relations. 
The relations in the $K$- and $V$-bands as well as in the $V,(V-I)$ 
Wesenheit index are shown in Figs.\ref{Fig.logPMk}-\ref{Fig.logPW} with the
Milky Way relations overplotted.

The dispersions of the Milky Way and LMC samples around the ridge line
PL relations given in Tab.\ref{tab.PLrelations}
are very similar for all the bands, and we note that there
is not a big difference in the dispersion as a function of photometric
band. The $J$ and $K$-band values are only marginally smaller than those
for the optical bands and the \Wvi index performs only
slightly better than the $V$-band. If however we consider the LMC Cepheids to 
be at the same distance we can determine apparent PL relations without
involving the individual distances from the IRSB method. In this case the 
dispersion reduces in \Wvi from 0.24 to 0.15~mag, in $V$ from 0.26 to
0.19~mag and in the $K$-band the value even decreases from 0.21 to 0.09~mag. 
Removing only two stars from the \Wvi relation further reduces that
disperion to 0.09~mag illustrating that the uncertainty on these numbers
are quite large due to the small sample size.  The fact that the
dispersion can be reduced so much by adopting a common distance to all
the stars suggests that the observed dispersion in the PL relations is 
dominated by errors in the distance moduli rather than intrinsic 
dispersion in the luminosities or reddening errors and the standard 
error on the individual distance moduli is about 0.2~mag.

The dispersion of the LMC and MW samples are also very similar suggesting
that the LMC data quality is equal to that for the Milky Way
sample.  Barnes et al. (\cite{Barnes05b}) performed a careful
Bayesian analysis of the data presented in Storm et al.  (\cite{Storm04})
and found that the formal error estimates from the regression fitting,
which are also the errors tabulated here in Tab.\ref{tab.LMCresults},
were systematically underestimated. On average these errors should
be multiplied by a factor of 3.4 to be in agremeent with the Bayesian
error estimates. In fact the error estimates for the distance moduli
presented in Tab.\ref{tab.LMCresults} have a mean value of 0.051~mag,
which results in a revised mean error estimate of 0.17~mag,  very
similar to the standard error of 0.2~mag estimated above from the
$K$-band PL relation.

\subsection{The effect of metallicity}
\label{sec.res_metal}

\begin{figure*}
\centering
\includegraphics[width=16cm]{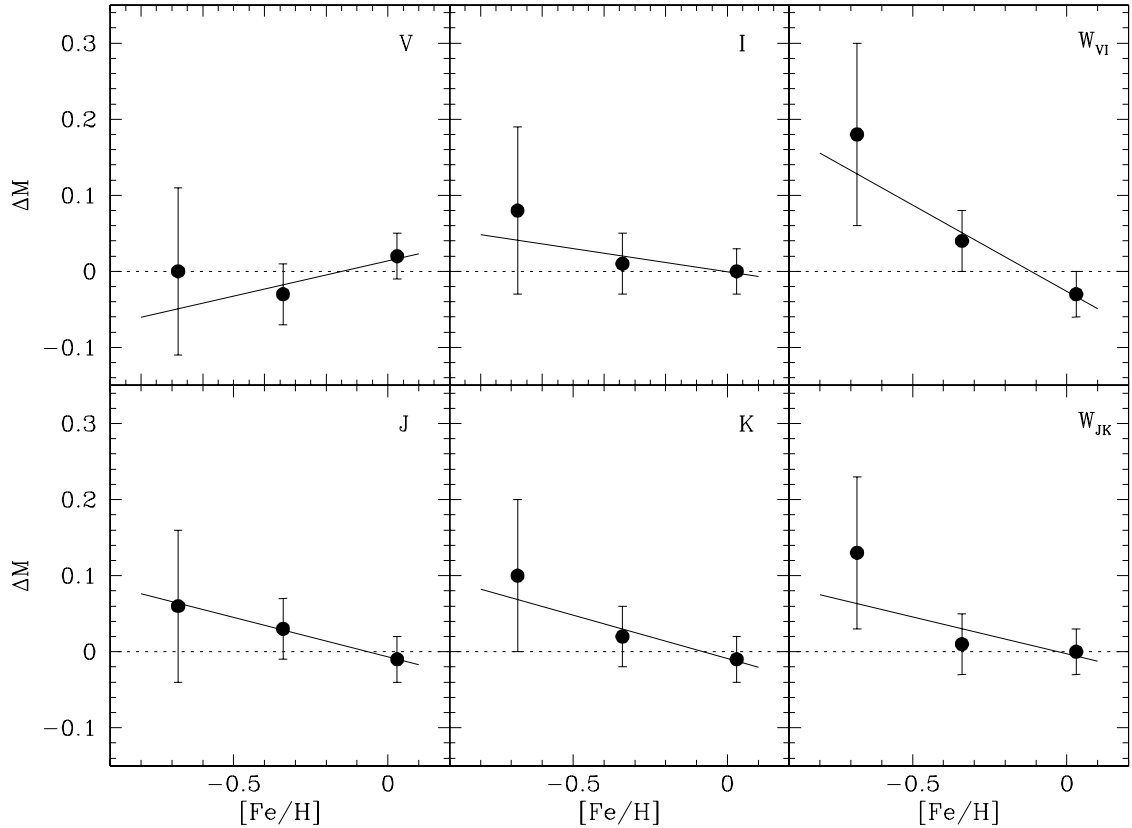}
\caption{\label{fig.FeHdM} The zero point offsets, $\Delta M =
M - M_{\mbox{\scriptsize comb}}$, for each 
band with respect to the reference PL relations from Paper I as a 
function of sample metallicity. The full lines show the weighted 
linear regression fit to the data. It can be seen that especially in the
limited but for extragalactic distance determination most important
metallicity range from LMC to solar abundance, the zero point offsets
are in general indistinguishable from zero. }
\end{figure*}

\begin{table}
\caption{\label{tab.FeHdM} The offsets of each metallicity sample
(Milky Way, LMC and SMC) with respect to the reference PL relations
from Paper I, i.e. assuming a fixed PL relation slope irrespective 
of metallicity. The last column gives the slope, $\gamma$, of the 
weighted linear regression fit to the data.}
\begin{tabular}{r c c c c}
\hline\hline
\multicolumn{1}{c}{Band} &
\multicolumn{1}{c}{$\Delta b$(MW)} & \multicolumn{1}{c}{$\Delta b$(LMC)} &
\multicolumn{1}{c}{$\Delta b$(SMC)} & \multicolumn{1}{c}{$\gamma$}\\
  & \multicolumn{1}{c}{(mag)} & \multicolumn{1}{c}{(mag)} &
\multicolumn{1}{c}{(mag)} & \multicolumn{1}{c}{(\magdex)} \\
\hline
 \Mv & $+0.02$ & $-0.03$ & $-0.00$ & $+0.09$\\
 \Mi & $-0.00$ & $+0.01$ & $+0.08$ & $-0.06$\\
\Wvi & $-0.03$ & $+0.04$ & $+0.18$ & $-0.23$\\
 \Mj & $-0.01$ & $+0.03$ & $+0.06$ & $-0.10$\\
 \Mk & $-0.01$ & $+0.02$ & $+0.10$ & $-0.11$ \\
\Wjk & $-0.00$ & $+0.01$ & $+0.13$ & $-0.10$\\
\hline
$\sigma$ & $\pm0.03$ & $\pm0.04$ & $\pm0.11$ & $\pm0.10$ \\
\hline
\end{tabular}
\end{table}

The LMC Cepheid PL relations in Tab.\ref{tab.PLrelations} can
be compared directly to the relations derived in Paper I for
the Milky Way sample which have been tabulated for convenience in
Tab.\ref{tab.PLrelations}. The slopes of the relations are in excellent
agreement in the case of the near-infrared $J$ and $K$ bands. In the $V$
band and in the Wesenheit indices, the agreement is slightly worse 
but well within one $\sigma$, whereas the $I$ band slopes differ at 
the level of about one $\sigma$.  Since the slopes
of the PL relations for both LMC and Milky Way Cepheids are very well
constrained from our samples, we conclude that our results present strong
evidence that the slope of the Cepheid PL relation, particularly in the
near-infrared $J$ and $K$ bands, is identical for the solar metallicity Milky
Way and more metal-poor LMC samples. These PL relations are thus 
independent of metallicity in the range from solar to LMC metallicity.  
The universality of the PL relation slopes appears to be confirmed 
in this metallicity range.

We can further extend the metallicity baseline by including the 5 SMC
Cepheids in the sample. Given the low number of Cepheids and the narrow
range of periods for these stars, we cannot constrain the slope of the
PL relation at this metallicty, but we can constrain the zero point
offset. In Fig.\ref{Fig.logPMkMC} we have overplotted the SMC Cepheids
on the LMC Cepheids for the $K$-band and obviously there is good agreement.
Considering the excellent agreement between the LMC and the Milky Way
from Tab.\ref{tab.PLrelations} this further supports
the universality of the zero point of the $K$-band relation.

We can quantify any offsets in the PL relation magnitude zero points 
as a function of metallicity by comparing the zero point offset for
each of the three samples with the reference PL relations determined in
Paper I on the basis of all available Cepheids. For each band and
sample we computed the mean magnitude offsets,
tabulated them in Tab.\ref{tab.FeHdM} and plotted them as a
function of metallicity in Fig.\ref{fig.FeHdM}. For each band we have
fit the weighted least square regression line to the magnitude offset as
a function of metallicity. The resulting
metallicity effect slopes have been tabulated in Tab.\ref{tab.FeHdM}. 
We estimate the error on the slopes to be of the order 0.10~\magdex 
based on the error bars on the individual points and the fact that 
the baseline is smaller than one dex. From the Fig.\ref{fig.FeHdM} and
Tab.\ref{tab.FeHdM} it can be seen that a zero metallicity effect
{\em cannot} be ruled out by the data in most cases but that the effect 
is significant and the strongest for the \Wvi index. For extra-galactic distance
determination most Cepheid samples will have metallicities in the range
from LMC to solar and the zero-point offsets in this range is clearly
very small in all the bands.

The only bands for which we have excellent agreement between the PL
relation slopes are the $J$ and $K$ bands, and the zero-point offsets
in both of these bands are small so they each form an excellent basis for
a standard candle. In the $V$ band and the \Wjk index
the PL relation slopes are marginally in agreement and also for these
two bands the zero-point offsets are small. The $I$ band and the \Wvi
index have significant differences in the PL slopes. The \Wvi index
 zero-point is also significantly more metallicity dependent than is
the case for the other bands making this index inferior to the $J$ and
$K$-bands as a standard candle.

  We note that we have assumed a simple linear metallicity dependence,
but it is of course entirely possible that the dependence, if present, has
a more complex functional form.

We emphasize that the metallicity effect on both the slopes and zero
points discussed here is {\em entirely independent of the choice of 
the $p$-factor relation}, whereas the absolute values of the slopes and
zero points as well as the resulting LMC and SMC distances do depend 
on the adopted $p$-factor relation. 



\clearpage 

\section{Discussion}
\label{sec.discussion}

We can now compare the slopes of our LMC PL relations with independent
measures from apparant magnitudes versus $\log (P)$.  
In the near-infrared we use the
relations from Persson et al. (\cite{Persson04}) which is based on
92 LMC Cepheids with excellent $J$ and $K$ light curves. They found
slopes of $-3.261\pm 0.042$ and $-3.153\pm 0.051$ in $K$ and $J$
respectively, in excellent agreement with our values of $-3.28\pm0.09$ 
and $-3.22\pm0.09$.  In the optical we can compare to the 
relations from the OGLE project (Udalski \cite{Udalski00}) and 
we similarly find very good agreement. They find in the $V$ and $I$-bands 
slopes of $-2.775\pm0.031$ and $-2.977\pm0.021$, 
to be compared with our values of $-2.78\pm0.11$ and $-3.02\pm0.10$ 
respectively. The fact that we reproduce the slopes of the 
PL relations in the LMC so well strongly supports our empirical
calibration of the $p$-factor used with the IRSB method and confirms
our earlier results (Gieren et al. \cite{Gieren05a}) based on a much
smaller sample of LMC Cepheids.

The LMC distance modulus of $18.45\pm0.04$ is only slightly shorter than the 
"canonical" value of $18.50\pm0.10$ as adopted by 
Freedman et al. (\cite{HSTkey}) and by the ARAUCARIA project 
(Gieren et al. \cite{Gieren05b}, Pietrzy\'nski et al.
\cite{Pietrzynski10a}), and slightly longer than the value of
$18.39\pm0.01(\mbox{random})\pm0.07(\mbox{systematic})$ suggested by Freedman and Madore (\cite{FM10}) in
their recent review. It is thus in good agreement with most recent results, 
as would be expected considering that we (see Paper I) have 
calibrated the IRSB method to match the distances to nine Milky Way 
Cepheids with direct parallaxes from Benedict et al.
(\cite{Benedict07}). These distances are accurate to 4-10\%,
which seems the largest single remaining systematic
uncertainty in the IRSB method assuming that the $p$-factor relation
has been accurately established from our new and improved constraints
(which is supported by the arguments given above). The total systematic
uncertainty of our present determination of the LMC distance is difficult
to quantify, but expected to be about $\pm$ 5\%, which is better than
most other techniques which have been used to measure the LMC distance
over the years. An exciting possibility to constrain the IRSB method
even further, in a very direct way, will be the comparison of the
IRSB-determined distance to the LMC Cepheid OGLE-LMC-CEP0227 which
is a member in a double-lined eclipsing binary system (Pietrzynski et
al. \cite{Pietrzynski10b}). Work is underway to determine the distance
to this binary from orbital analysis.

  In Sec.\ref{sec.res_metal} we compared the slopes of the PL relations 
from the Milky Way and LMC samples finding insignificant differences in
the near-IR bands and small (but possibly also nonsignificant) differences 
in the optical bands. This result is in good agreement with
the findings by Gieren et al. (\cite{Gieren05a}) but at odds with
earlier findings by us (Fouqu\'e, Storm and Gieren, \cite{Fouque03},
Storm et al. \cite{Storm04}) where we found that the slope of the 
galactic relations were significantly steeper than those
in the LMC and SMC. The reason for this difference was the inappropriate
$p$-factor relation used in the earlier work for determining the Milky Way 
PL relations. Note that in the present work the choice of the
$p$-factor relation has no bearing on the difference in slope between
the Milky Way and LMC PL relations as we are now applying the method to
both samples of Cepheids, this was not possible previously as the
necessary data for the LMC stars was not yet available.
Sandage, Tammann, and Reindel (\cite{STR04}) also argued that there 
is a significant difference in slope between the Milky Way and 
LMC PL relations in the optical bands, the Milky Way relation being 
steeper. This conclusion was largely based on these older IRSB results, 
as well as on Cepheids in OB associations from Feast (\cite{Feast99}).
As we show in Paper I the revised IRSB distances are,
apart from a small zero point offset, in very good agreement with the latest
results on open cluster Cepheids from Turner (\cite{Turner2010}) and
does not exhibit a period dependence.

We do not support the conclusion of a strong effect
of metallicity on the slope of optical $(V,I)$ Cepheid PL relations as
reached, for example, by Tammann et al. (\cite{Tammann11}) or Storm et
al. (\cite{Storm04}), in fact we find that for these bands the 
Milky Way PL relations are slightly shallower than the LMC relations, 
only for the \Wjk index we find that the Milky Way relation is steeper
than the LMC relation.

  In Tab.\ref{tab.FeHdM} we summarized the PL relation zero point
variation as a function of metallicity. We have made the simplest
assumption of a linear relation and found the slopes of the zero point
offset versus metallicity relations, $\gamma$. Clearly the effects in
the near-IR are small and even in the $V$ band the effect appears to be
small, albeit with the opposite sign as for the other bands. The largest
effect we find is in the \Wvi index and our value of
$\gamma(\Wvi)=-0.23\pm0.10$~\magdex is in excellent agreement with the
value of $-0.24\pm0.05$~\magdex found by Sakai et al. (\cite{Sakai04}) 
and other measurements which have been adopted in the recent review 
paper by Freedman and Madore (\cite{FM10}). This result is slightly at
odds with the findings of Bono et al. (\cite{Bono10}) who on the basis
of data on 48 external galaxies finds no significant metallicity 
effect on neither slope nor zero point for the Wesenheits indices 
\Wvi and \Wjk. At the same time we do agree with them
that the PL relation slopes are less affected in the $J$ and $K$-bands 
and more affected in the optical $V$ and $I$ bands and we also agree on
the most likely sign of the effect namely that metal-rich PL relations
are shallower than metal-poor ones.

It is interesting to note that the most significant metallicity
effect is found for the \Wvi index and that this effect
can largely be attributed to the colour difference between Milky Way and
LMC Cepheids as found by Sandage, Tamman and Reindl (\cite{STR04}). They
compared period-colour relations for both Milky Way and LMC Cepheids
and found that the LMC Cepheids for a given period are bluer by about
0.05~mag. Feeding this into the Wesenheits index ($\Wvi = V - 2.54 ( V
- I)$), 
we find an offset of $0.00 - 2.54 \times -0.05 = 0.13$~mag in the case
where the offset is fully in the $I$-band and 
$-0.05 - 2.54 \times -0.05 = 0.08$~mag in the case where the effect is 
fully in the $V$-band.  Disregarding the subtle effects of
differences in the slopes of the period-colour relations, these results
are comparable to the offset of 0.09~mag, metal-poor Cepheids being
fainter, which we find between the Milky Way and LMC \Wvi relations in 
Tab.\ref{tab.PLrelations} at a period of 10~days. 


The emerging conclusion based on our data and analysis
is that for accurate distance measurements to
galaxies the $K$-band Cepheid PL relation is the best suited tool:
it is metallicity-independent both regarding the slope and the zero
point, it is very insensitive to reddening, and it has a
smaller intrinsic dispersion than any optical PL relation. It is likely,
as indicated in recent work from Spitzer data, that mid-infrared Cepheid
PL relations are even superior to their near-infrared relations because
of their even lower sensitivity to reddening, and lower intrinsic
dispersions (Madore et al. \cite{Madore09}).  Yet, their dependence on
metallicity has still to be investigated and they cannot be exploited
from the ground making them exceedingly expensive to use.

\section{Conclusions and Summary}
\label{sec.summary} We have obtained new and very accurate radial
velocity curves for 22 LMC Cepheids thus expanding the sample of LMC
Cepheids to 36 for which we can apply the IRSB distance analysis.
We have applied the newly calibrated IRSB technique of Paper I 
to these 36 LMC Cepheids as well as to 5 SMC Cepheids. 
The IRSB analysis yields individual distances 
from which we calculate absolute magnitudes in optical
and near-infrared bands.  These magnitudes define tight Period-Luminosity
relations in the $V, I, J, K$ bands and in the Wesenheit indices.  

We show that the PL relations are in excellent agreement with the
observed apparent magnitude versus $\log (P)$ relations in both the
optical and near-IR bands lending strong support to the empirical 
calibration of the $p$-factor relation in Paper I.

By comparing the LMC Cepheid PL relations to their Milky Way counterparts 
reported in Paper I that were established with exactly the same precepts, 
we find
practically identical Milky Way and LMC PL relation slopes in the near-infrared
bands, arguing for the universality of the Cepheid PL relation in this
spectral range. The zero points exhibit a slight metallicity effect,
$\gamma(\Mk)=-0.10\pm0.10$~\magdex in the sense that metal-poor Cepheids
are fainter than metal-rich Cepheids. If we restrict ourselves only to the
metallicity range between solar and LMC, our results are consistent with
universal PL relations (in both slope and zero point) in the near-infrared
$J$ and $K$ bands.

In the optical bands, we find that the slopes depend weakly
on metallicity, the Milky Way slopes being slightly shallower than the
LMC slopes, but this difference might not be significant.

The optical PL relation zero points exhibit a
metallicity effect of similar size as in the near-IR albeit the sign is
opposite in the $V$-band. The \Wvi index shows the strongest zero point
effect on metallicity where we find $\gamma(\Wvi)=-0.23\pm0.10$~\magdex.
We stress that the zero-point offsets are based on a rather long
metallicity baseline ranging from SMC to solar metallicity.

Our direct distances to the LMC Cepheids leads to a true LMC distance 
modulus of $18.45 \pm 0.04$~mag, with an estimated systematic uncertainty 
of 5\% which mainly comes from the uncertainty on the HST parallaxes of nine
Milky Way Cepheids that have been used to define the absolute zero point
of the IRSB technique. Our distances to the SMC Cepheids leads to an SMC
distance of $18.92\pm0.14$.

Both the \Wvi metallicity effect and the LMC distance which we find
are in agreement with the latest values adopted by 
Freedman and Madore (\cite{FM10}) in their recent review on the Hubble 
constant. 

Considering the significant metallicity effect on the \Wvi index, 
not only on the zero-point but most likely also on the slope, we
argue that the best standard candle is presently provided by the PL
relation in the $K$-band as it is metallicity insensitive, reddening
insensitive and exhibits the lowest intrinsic scatter.

\acknowledgements{We thank Roeland van der Marel for providing his code
for computing the LMC distance correction for the stars. A great thanks
is due to the La Silla support staff and in particular to the 
service mode team which managed so well to cover the phase gaps left
after the first visitor mode run.
We gratefully acknowledge financial support for this work from the
Chilean Center for Astrophysics FONDAP 15010003, and from the BASAL
Centro de Astrofisica y Tecnologias Afines (CATA) PFB-06/2007.
}

\end{document}